\documentclass[12pt,a4]{article}
\usepackage{amsmath}
\usepackage{amssymb}
\usepackage{epsfig}
\usepackage{subfigure}
\usepackage{hyperref}
\usepackage{mathrsfs}

\begin{document}
\pagestyle{plain}
\begin{titlepage}
\begin{center}
\hfill    CERN-PH-TH/2012-169\\

\vskip 1cm


{\large \bf {Invariants and Flavour in the General  Two-Higgs Doublet Model}}

\vskip 1cm

F. J. Botella  $^a$ \footnote{fbotella@uv.es}, 
G. C. Branco  $^b$ \footnote{gustavo.branco@cern.ch and 
gbranco@ist.utl.pt},  and M. N.
Rebelo $^c$ \footnote{margarida.rebelo@cern.ch and
rebelo@ist.utl.pt}

\end{center}

\vspace{1.0cm}

\noindent
{\it $^a$ Departament de F\' \i sica Te\`orica and IFIC,
Universitat de Val\`encia-CSIC, E-46100, Burjassot, Spain.} \\
{\it $^b$ Departamento de F\'\i sica and Centro de F\' \i sica Te\' orica
de Part\' \i culas (CFTP),
Instituto Superior T\' ecnico, Av. Rovisco Pais, P-1049-001 Lisboa,
Portugal.} \\
{\it  $^c$ Universidade T\' ecnica de Lisboa, Centro de F\' \i sica Te\' orica
de Part\' \i culas (CFTP), Instituto Superior T\' ecnico, Av. Rovisco Pais, 
P-1049-001 Lisboa, Portugal.} \\

\begin{abstract}
The flavour structure of the general Two Higgs Doublet Model (2HDM) is 
analysed and a detailed study of the parameter space is presented, showing
that flavour mixing in the 2HDM can be parametrized by various
unitary matrices which arise from the misalignment in flavour space
between pairs of various Hermitian flavour matrices which can be constructed
within the model. This is entirely analogous to the generation of the CKM matrix
in the Standard Model (SM). We construct weak basis invariants which can give
insight into the physical implications of any flavour model, written in an
arbitrary weak basis (WB) in the context of 2HDM. We apply this technique to
two special cases, models with MFV and models with NNI structures. 
In both cases non-trivial 
CP-odd WB invariants arise in a  mass power order much smaller than what 
one encounters in the SM, which can have important implications for baryogenesis 
in the framework of the  general 2HDM.
\end{abstract}

\end{titlepage}

\newpage

\section{Introduction}
The two Higgs doublet model (2HDM) \cite{review} is one of the
simplest extensions of the Standard Model (SM) and
arises in many models beyond the SM, including supersymmetric (SUSY) ones.
The 2HDM was first introduced by Lee \cite{Lee:1973iz}  with the aim
of achieving spontaneous CP violation \cite{Branco:1999fs} in the context of the SM,
at a time when only two incomplete fermion generations were known.
No extra symmetries are introduced in Lee's model and, as a result, 
the model has flavour-changing-neutral currents (FCNC) of arbitrary
strength at tree level. In order to avoid FCNC at tree level in the 2HDM,
a discrete $Z_2$ symmetry can be introduced \cite{Glashow:1976nt}, 
which guarantees Natural Flavour Conservation (NFC) in the scalar sector.
It was pointed 
out that in this case neither spontaneous \cite{Branco:1980sz} 
nor hard CP violation \cite{Weinberg:1976hu}
in the Higgs sector can be achieved, unless one introduces a third 
Higgs doublet. An alternative scenario is to break this $Z_2$ 
symmetry softly \cite{Branco:1985aq}.
In this paper we study the flavour content of the general 2HDM and
construct weak basis (WB) invariants which can give insight 
into the physical implications of any flavour model written in an arbitrary WB.
It should be stressed that even in models where each charge quark sector
receives mass contributions from only one Higgs, as it is the case in 
SUSY models, ``wrong" couplings are generated at higher orders
\cite{Hamzaoui:1998nu}, \cite{Babu:1999hn}, \cite{Isidori:2001fv}. 
Therefore, the present analysis maybe relevant also for models
with NFC at tree level.

At this stage it is worth recalling that in the 
presence of a flavour symmetry or an Ansatz,
the Yukawa couplings may contain texture zeros which arise only 
in a specific basis. In another WB  the Yukawa coupling matrices
change, the texture zeros may no longer be present but the physical content
of the model does not change. The great advantage of the WB invariants
stems from the fact that they can be evaluated in any WB. Furthermore,
we point out that the flavour structure in the 2HDM can be parametrized by 
various unitary matrices which are entirely analogous to the 
Cabibbo-Kobayashi-Maskawa (CKM) matrix of the SM. All the unitary flavour
mixing matrices of the 2HDM arise from the misalignment in flavour space of various
Hermitian matrices constructed in the framework of the 2HDM.
In order to illustrate the usefulness of these WB invariants, we apply them 
to the analysis of 2HDM which have Higgs mediated FCNC at tree level (HFCNC), 
but with their structure 
entirely defined \cite{Branco:1996bq} , \cite{Botella:2009pq}
in terms of $V_{CKM}$. It has been pointed out
that some of these models satisfy the hypotheses of Minimal Flavour Violation
\cite{D'Ambrosio:2002ex} (see also \cite{Chivukula:1987py}, \cite{Hall:1990ac},
\cite{Buras:2000dm}) .
The paper is organized as follows: in the next section, we settle the notation and 
analyse the flavour parameter space of the general 2HDM, explaining how
the various unitary flavour mixing matrices are generated. In section 3 
we display how the various Yukawa couplings transform under WB transformations,
construct various WB invariants and analyse their physical meaning.
Next we illustrate how WB invariants can be used to analyse specific 
flavour models, based on 2HDM. As examples we use the class of models
named BGL  \cite{Branco:1996bq}, where there are FCNC at tree level but with 
their flavour structure controlled by $V_{CKM}$,  and a model with 
nearest-neighbour-interaction (NNI) pattern for the quark mass matrices in
the framework of a 2HDM \cite{Branco:2010tx}. 
Finally in section 4 we present our Conclusions.

\section{The Two Higgs Doublet Parameter Space}
We consider the extension of the SM consisting of the addition
of two Higgs doublets (2HDM) with no additional symmetries.
This implies that each of the doublets $\Phi _{1}$,  $\Phi _{2}$
contributes to both up and down quark mass matrices, through
the Yukawa couplings:
\begin{eqnarray}
\mathcal{L}_{Y} \ = \  -\ \overline{Q_{L}^{0}}\ \Gamma _{1}\Phi _{1}d_{R}^{0}-
\overline{Q_{L}^{0}}\ \Gamma _{2}\Phi _{2}d_{R}^{0}-\overline{Q_{L}^{0}}\
\Delta _{1}\tilde{\Phi }_{1}u_{R}^{0}-\overline{Q_{L}^{0}}\ \Delta _{2}
\tilde{\Phi }_{2}u_{R}^{0} + \text{h.c.} 
\label{LY}
\end{eqnarray}
where we have used standard notation. The interactions of the neutral Higgs 
with the quarks, obtained from Eq.(\ref{LY})  are given by:
\begin{eqnarray}
{\mathcal L}_Y (\mbox{neutral})& = & - \overline{d_L^0} \frac{1}{v}\,
[M_d H^0 + N_d^0 R + i N_d^0 I]\, d_R^0  + \nonumber \\
&-& \overline{{u}_{L}^{0}} \frac{1}{v}\, [M_u H^0 + N_u^0 R - i N_u^0 I] \,
u_R^{0} + \text{h.c.}\ ,
\label{rep}
\end{eqnarray}
with  $v \equiv \sqrt{v_1^2 + v_2^2}$, and $H^0$, $R$  orthogonal combinations 
of the fields  $\rho_j$, given by $ \phi^0_j =  \frac{e^{i \theta_j}}{\sqrt{2}} 
(v_j + \rho_j + i \eta_j)$, where $H^0$ is defined so that its couplings
are proportional to the mass matrices. In an analogous way, $I$ is a
linear combination of $\eta_{j}$ orthogonal to the neutral Goldstone boson. 
The quark mass matrices $M_d$ and $M_u$ and the matrices $N_d^0$ and $N_u^0$ 
are given by:
\begin{eqnarray}
M_{d}=\frac{1}{\sqrt{2}}(v _{1}\Gamma _{1}+v _{2}e^{i\theta
}\Gamma _{2})\ ,\qquad 
M_{u}=\frac{1}{\sqrt{2}}(v _{1}\Delta
_{1}+v _{2}e^{-i\theta}\Delta _{2})\ , \nonumber \\
N_d^0 = \frac{v_2 }{\sqrt{2}} \Gamma_1  - \frac{v_1 }{\sqrt{2}} 
e^{i \theta} \Gamma_2\ , \qquad
N_u^0 = \frac{v_2 }{\sqrt{2}} \Delta_1  - \frac{v_1 }
{\sqrt{2}} e^{-i \theta} 
\Delta_2\ , 
\label{quatro}
\end{eqnarray}
where  $\theta$ denotes the relative phase
of the vevs of the neutral components of $\Phi_i$.
The four matrices of Eq.(\ref{quatro}) are written in an arbitrary weak-basis (WB).
It is well known that one can make a WB transformation defined by:
\begin{eqnarray}
d_L^0= W_L \ {d_L^0}^\prime, \qquad d_R^0= W_R^d \ {d_R^0}^\prime, \qquad
u_L^0= W_L \ {u_L^0}^\prime, \qquad u_R^0= W_R^u \ {u_R^0}^\prime
\end{eqnarray}
without physical implications. Under these WB transformations, the matrices of
Eq.~(\ref{quatro}) transform as:
\begin{eqnarray}
M_d \rightarrow M_d^\prime  = W_L^\dagger M_d W_R^d,  \qquad
M_u \rightarrow M_u^\prime  = W_L^\dagger M_u W_R^u, \nonumber \\
N_d^0  \rightarrow {N_d^0} ^\prime  = W_L^\dagger N_d^0  W_R^d,  \qquad
N_u^0 \rightarrow {N_u^0}^\prime  = W_L^\dagger N_u^0 W_R^u
\label{wbtra}
\end{eqnarray}
In order to analyse the physical content of the above four matrices,
one may choose, without loss of generality, a weak-basis where $M_u$ is 
diagonal real, while $M_d$ is a Hermitian matrix with only one rephasing 
invariant phase given by $\varphi = \arg[{(M_d)}_{12} {(M_d)}_{23} {(M_d)}_{31} ]$.
The six real parameters in $M_d$, together with   $\varphi $
and the up quark masses $m_u$, $m_c$, $m_t$, total the ten parameters contained 
in the flavour sector of the SM, seen in a weak basis. In the quark mass eigenstate basis
these appear as the six quark masses and the four parameters characterizing $V_{CKM}$. 
In the above described WB, the matrices $N_d^0$, $N_u^0$ are in general 
complex arbitrary $3 \times 3$ matrices, each one containing nine physical phases.
Note that we have considered the general 2HDM with no flavour 
symmetries introduced. 

In the presence of flavour symmetries and/or 
texture zeros, the number of parameters in $N_d^0$, $N_u^0$ can be drastically
reduced. Flavour symmetries (FS) are introduced in a specific WB, with the choice
dictated by the FS representation assumed for the fermions and Higgs doublets.
Similarly, texture zeros imply the choice of a particular WB. In view of this freedom of
choice of WB, it is very useful to express the physical content of  $M_d$,  $M_u$,
 $N_d^0$,  $N_u^0$ in terms of WB invariants.

We shall construct these invariants and analyse their physical content in section 3.

It is useful to see how the parameters of $N_d^0$,  $N_u^0$  appear when one 
parametrizes  $N_d^0$,  $N_u^0$ through unitary matrices. It can be readily seen
that, without loss of generality, one can write:
\begin{equation}
N_d^0 = K_L \ \hat{V}_L^{N_d}\   D^{N_d} \ \overline{K} \ (\hat{V}_R^{N_d})^\dagger \ K_R^\dagger 
\end{equation} 
where $K_L$, $K_R$ are diagonal unitary matrices of the form:
\begin{equation}
K_{L,R} = \mbox{diag}[ 1, \exp(i \varphi_{1L,R}), \exp(i \varphi_{2L,R}) ]
\end{equation} 
while $\hat{V}_{L,R}^{N_d}$ are unitary matrices with one physical 
non factorizable phase each, 
analogous to $V_{CKM}$. Finally one has:
\begin{equation}
\overline{K} = \mbox{diag}[ \exp(i\sigma_1), \exp(i \sigma_2), \exp(i \sigma_3) ]
\end{equation} 
and $  D^{N_d}$ stands for a real diagonal matrix.
The explicit counting of parameters is:
\begin{eqnarray}
\mbox{phases:} \qquad 2 (K_L) + 2 (K_R) + 1 (\hat{V}_L^{N_d} )+ 1 (\hat{V}_R^{N_d}) + 
3 (\overline{K}) = 9 \nonumber \\
\mbox{real parameters} \qquad 3 (\hat{V}_L^{N_d}) + 3 (\hat{V}_R^{N_d}) + 3 (D^{N_d}) = 9
\nonumber
\end{eqnarray}
for each one of the matrices $N_d^0$,  $N_u^0$.

\section{Weak Basis Invariants}
\subsection{The General Case}
The four matrices, $M_d$, $M_u$, $N_d^0$,  $N_u^0$ fully characterize 
the flavour sector of the 2HDM in the sense that they encode the breaking of
the large flavour symmetry present in the gauge sector of the theory. The above 
four flavour matrices contain a large redundancy of parameters which results from 
the fact that under a WB transformation  $M_d$, $M_u$, $N_d^0$,  $N_u^0$
change transforming as indicated by Eq.~(\ref{wbtra})  without altering their 
physical content. Different Lagrangians related to each other by WB transformations
describe the same physics. In view of the above redundancy, it is of great 
interest to construct WB invariants which can be very useful in the analysis of 
the physical content of the flavour sector of a given model. For example,
in the context of the SM, it has been shown \cite{Branco:1987mj}  that from the 
four WB  invariants tr$ (H_u\ H_d)$,  tr$(H_u\ H^2_d)$,  tr$(H^2_u\ H_d)$, 
 tr$ (H^2_u\ H^2_d)$, where $H_{d,u} \equiv (M_{d,u} M^\dagger _{d,u})$, one
can construct the full $V_{CKM}$, with only a two-fold ambiguity in the sign
of Im$Q$, where $Q$ stands for a rephasing invariant quartet of $V_{CKM}$,
defined by $Q_{\alpha i \beta j} \equiv V_{\alpha i}  V_{\beta j}  V^\ast_{\alpha j}  
V^\ast_{\beta i} $ ($\alpha \neq \beta$, $ i \neq j$). WB invariants are also very useful 
in the study of CP violation. In the context of the SM, it has been derived from first principles
 \cite{Bernabeu:1986fc} that the necessary and sufficient condition for CP 
invariance is the vanishing of the WB invariant:
\begin{eqnarray}
I_1^{CP} \equiv & \mbox{tr} \left[  H_u, H_d \right]^3  = 6i  (m^2_t - m^2_c)   
(m^2_t - m^2_u)   (m^2_c - m^2_u) \times \nonumber \\ 
& \times (m^2_b - m^2_s)   (m^2_b - m^2_d) (m^2_s - m^2_d)  \mbox{Im} Q_{uscb}
\label{eq31}
\end{eqnarray}
for three generations $I_1^{CP} $ is proportional to det$\left[  H_u, H_d \right]^3 $,
introduced in Ref.~\cite{Jarlskog:1985ht}.

In this section we use WB invariants to analyse the flavour structure
and CP violation in the general 2HDM. We shall apply here the same technique 
that was introduced in \cite{Bernabeu:1986fc} to the study  of CP violation in the
SM. This technique  was later generalized to many different scenarios, in particular to
the study of explicit CP violation in the scalar sector of multi-HDM prior to 
gauge symmetry breaking \cite{Branco:2005em} as well as CP violation in the scalar sector 
after this breaking \cite{Lavoura:1994fv}
 and also taking into account both the scalar and the 
fermionic sector \cite{Botella:1994cs}.   In Ref.~\cite{Dreiner:2007yz} CP violation in the 
supersymmetric case is analysed. WB invariants can also be built to 
study other important features of flavour models such as alignment and the pattern of fermion masses and mixing \cite{Branco:2011aa}. One can check the predictions of a flavour model
by comparing invariant quantities with their corresponding experimental
values. In Ref. \cite{Jenkins:2009dy}, the authors  classified 
all the invariants that can be built in a given theory, using the ring
of polynomials that are invariant under the action of a group.

From the transformation
properties of the flavour matrices  $M_d$, $M_u$, $N_d^0$,  $N_u^0$ given 
in Eq.~(\ref{quatro}), it is clear that one can build new WB invariants, which do not arise
in the SM, by evaluating traces of blocks of matrices involving
the up and down quark sector, like for example $M_\gamma N_\gamma^{0 \dagger}$ or
$N_\gamma^0 N_\gamma^{0 \dagger}$. We shall analyse the lowest WB invariants
and indicate some of the physical aspects of the 2HDM probed by each one 
of these invariants. For definiteness let us consider the WB invariant 
tr($M_d N_d^{0 \dagger}$) and note that its physical significance becomes
transparent in the WB where $M_d$ is diagonal, real, since in this basis the
matrix $N_d^0$ already coincides with  the couplings to the  physical quarks. 
In this basis one has:
\begin{equation}
I_1 \equiv \mbox{tr} (M_d N_d^{0 \dagger}) = m_d(N_d^\ast)_{11} +
m_s(N_d^\ast)_{22} + m_b(N_b^\ast)_{33}
\label{nnn}
\end{equation}
We denote $N_d$, the matrix $N_d^0$ in the basis where it couples to the 
physical quarks. This invariant is not sensitive to  Higgs-mediated FCNC, but Im($I_1$)
is specially important, since it probes the phases of $(N_d)_{jj}$ which 
contribute to the electric dipole moment of down-type quarks. Obviously,
one can construct an analogous invariant for the up-quark sector, namely
tr($M_u N_u^{0 \dagger}$). Let us now consider a WB invariant
which is sensitive to the off-diagonal elements of $N_d$, namely:
 \begin{eqnarray}
I_2 \equiv \mbox{tr} \left[ M_d N_d^{0 \dagger}, M_d M_d^ \dagger \right]^2 
 = -2 m_d m_s (m_s^2 - m_d^2)^2 (N_d^\ast)_{12} (N_d^\ast)_{21} -
\nonumber \\
- 2 m_d m_b (m_b^2 - m_d^2)^2 (N_d^\ast)_{13} (N_d^\ast)_{31}  -
2 m_s m_b (m_b^2 - m_s^2)^2 (N_d^\ast)_{23} (N_d^\ast)_{32} ,
\end{eqnarray}
where we have kept the notation used in Eq.~(\ref{nnn}), having evaluated $I_2$
in the WB where $M_d$ is real and diagonal. It is well known that $I_1^{CP} $ given
in Eq.~(\ref{eq31})
measures the strength of CP violation arising from weak charged currents
with the appearance of a non-trivial quark mixing matrix 
$V_{CKM} \equiv  U^\dagger_{uL}  U_{dL}$ reflecting the fact that 
$ U_{dL} \neq  U_{uL}$, i.e. the misalignment of the matrices
$H_d$, $H_u$ in flavour space. In an entirely analogous way, one can 
construct the invariant:
\begin{eqnarray}
I_2^{CP} \equiv & \mbox{tr} \left[  H_u, H_{N_d^0} \right]^3  = 
6i \Delta_u \Delta_{N_d}  \mbox{Im} Q_2
\label{i2cp}
\end{eqnarray}
where $Q_2$ is a rephasing invariant quartet of
$V_2 \equiv  U^\dagger_{uL}  U_{N^0_{d}L}$, $\Delta_u \equiv 
 (m^2_t - m^2_c)   (m^2_t - m^2_u)   (m^2_c - m^2_u)$
and $\Delta_{N_d} $ is defined in analogy to  $\Delta_u$ but  refering
to the eigenvalues of  $  H_{N_d^0} \equiv N_d^0  N_d^{0 \dagger}$.
It is clear that $V_2$ reflects the misalignment of the matrices 
$H_u$,  $  H_{N_d^0} $ in flavour space. Similarly, one has the
invariant :
\begin{eqnarray}
I_3^{CP} \equiv & \mbox{tr} \left[  H_d,  H_{N_d^0}  \right]^3  = 
6i \Delta_d \Delta_{N_d}  \mbox{Im} Q_3
\label{i3cp}
\end{eqnarray}
where $Q_3$ is a rephasing invariant quartet of $V_3 \equiv  U^\dagger_{dL}  U_{N^0_{d}L}$.
In an entirely analogous way, one can also construct the invariants:
\begin{eqnarray}
I_4^{CP} \equiv  \mbox{tr} \left[  H_u,  H_{N_u^0} \right]^3 ; \quad
I_5^{CP} \equiv  \mbox{tr} \left[  H_d,  H_{N_u^0} \right]^3 ; \quad
	I_6^{CP} \equiv  \mbox{tr} \left[   H_{N_d^0},  H_{N_u^0} \right]^3 
\label{icps}
\end{eqnarray}
which are proportional to the imaginary parts of the invariant quartets of 
$ U^\dagger_{uL}   U_{N^0_{u}L}$,  $U^\dagger_{dL}   U_{N^0_{u}L}$ and
$  { U^\dagger_{N^0_{d}L}}   U_{N^0_{u}L}$ respectively.  So far, we have only considered
invariants which are sensitive to left-handed mixings. One can construct 
analogous invariants which are sensitive to right-handed mixings, like:
\begin{eqnarray}
I_7^{CP} \equiv  \mbox{tr} \left[  H_d^\prime,  H^\prime_{N_d^0} \right]^3  = 
6i \Delta_d \Delta_{N_d}  \mbox{Im} Q_7
\label{i7cp}
\end{eqnarray}
where $H_d^\prime \equiv M^\dagger _d M_d $,
$  H^\prime_{N_d^0} \equiv N_d^{0 \dagger} N_d^0$
and $Q_7$ is a rephasing invariant quartet of  $U_{dR}  U_{N_d^0R}^\dagger$.
Obviously, one can construct analogous invariants with the up  sector, namely  
$I_8^{CP} \equiv  \mbox{tr} \left[  H_u^\prime,  H^\prime_{N_u^0} \right]^3$.

\subsection{The Minimal Flavour Violation Case}
The invariants considered in the general 2HDM can obviously be applied to any
flavour model where the matrices $\Gamma _{1}$, $\Gamma _{2}$, $\Delta _{1}$
and $\Delta _{2}$ have specific flavour structures (e.g. texture zeros) resulting,
for example, from a flavour symmetry introduced in the Lagrangian. As we have
seen, in the general 2HDM, the flavour structure of $N_d^0$, $N_u^0$ is arbitrary, 
which may lead to dangerous Higgs mediated FCNC, unless some natural 
suppression mechanism is found. Some time ago a class of models
was constructed by Branco, Grimus and Lavoura (BGL) \cite{Branco:1996bq} 
where HFCNC  are present at tree level with their structure  entirely controlled 
by $V_{CKM}$ with no other new flavour parameters. The class of models 
considered in Ref.~\cite{Branco:1996bq} 
are entirely natural since their remarkable features result from a symmetry 
imposed on the Lagrangian. These models were generalized  and their MFV character
was analysed in Ref.\cite{Botella:2009pq}. An extension to 
the leptonic sector was proposed \cite{Botella:2011ne}, with the r\^ ole of $V_{CKM}$
replaced by the Pontecorvo-Maki-Nakagawa-Sakata matrix denoted $V_{PMNS}$.
The MFV hypothesis requires that the flavour structure of physics beyond the
SM should only depend on $V_{CKM}$ entries, quark masses and, in the case 
of 2HDM, on the ratio $v_1/v_2$ of Higgs vevs, with the corresponding analogue
for the leptonic sector. The MFV as defined in \cite{D'Ambrosio:2002ex}
also requires that the breaking of the flavour symmetry be dominated by
the top Yukawa couplings. In the context of the 2HDM this leads to the requirement 
that the new physics beyond the SM should be suppressed by the third
row of $V_{CKM}$ in order to comply with all the criteria introduced in the
original paper where the definition of the Minimal Flavour Violation 
hypothesis was introduced  \cite{D'Ambrosio:2002ex}.

For definiteness let us consider the Yukawa couplings arising in a 
specific BGL model, which realizes the MFV hypothesis with HFCNC
only in the down sector:
\begin{eqnarray}
 \Gamma_1 & = & \left[\begin{array}{ccc}  
\times  & \times & \times \\
\times & \times &  \times \\
0 & 0 & 0 
\end{array}\right]; \qquad
 \Gamma_2   =  \left[\begin{array}{ccc}  
0 & 0 & 0  \\
0 & 0 & 0 \\
\times & \times &  \times 
\end{array}\right] \label{gam}\\
 \Delta_1  & = & \left[\begin{array}{ccc}  
\times  & \times & 0 \\
\times & \times &  0 \\
0 & 0 & 0 
\end{array}\right]; \qquad 
 \Delta_2   =  \left[\begin{array}{ccc}  
0  & 0 & 0 \\
0 & 0 &  0 \\
0 & 0 & \times
\end{array}\right] \label{del}
\end{eqnarray}
This BGL model realizes the MFV hypothesis in a natural way. It has been pointed out
\cite{Branco:1996bq} that there are six BGL models which correspond  to interchanges 
of rows in the matrices given above as well as choosing what sector (up or down)
has HFCNC which amounts to interchanging the matrices 
$ \Gamma_i$ with the matrices $ \Delta_i$. 
As previously emphasized, the specific texture of Eqs.~(\ref{gam}), (\ref{del}),
reflects a particular choice of WB. In the sequel, we  give WB-independent 
necessary and sufficient conditions for a set of Yukawa couplings 
$\Gamma_i$, $ \Delta_i$ written in an arbitrary WB to be of the BGL type,
implying the existence of a WB where these matrices can be cast in the form given above. \\

\noindent
{\bf \underline {Necessary and Sufficient Conditions for BGL}} \\

The following relations:
\begin{eqnarray}
 \Delta_1^\dagger   \Delta_2 =0; \quad 
  \Delta_1  \Delta_2^\dagger  =0;   \quad
\Gamma_1^\dagger  \Delta_2 =0; \quad
 \Gamma_2^\dagger  \Delta_1 =0
\label{gadel} 
\end{eqnarray}
are necessary and sufficient conditions for a set of Yukawa
matrices $\Gamma_i$, $ \Delta_i$ to be of the BGL type, with
Higgs mediated FCNC in the down sector. \\

\noindent
{\bf \underline {Proof} }\\
Note that the conditions of Eqs.~(\ref{gadel}) are WB independent, in the sense that
if a set of matrices  $\Gamma_i$, $ \Delta_i$ satisfy Eqs.~(\ref{gadel}) 
in a given  WB, they will satisfy them when written in any other WB. 
From  Eqs.~(\ref{gadel})  it follows that:
\begin{eqnarray}
\left[  \Delta_1   \Delta_1^\dagger,  \Delta_2   \Delta_2^\dagger \right] =0; \quad 
\left[ \Delta_1^\dagger \Delta_1, \Delta_2^\dagger \Delta_2 \right] =0 
\label{dddd}
\end{eqnarray}
From Eq.~(\ref{dddd}) one concludes that one can choose
a basis where both  $\Delta_1$ and  $\Delta_2$ are diagonal, real:
\begin{eqnarray}
\Delta_1= d_1  \equiv  \mbox{diag.} \left[
(d_1)_1,  (d_1)_2,  (d_1)_3 
\right]; \ \ 
\Delta_2= d_2  \equiv  \mbox{diag.} \left[  
(d_2)_1,  (d_2)_2,  (d_2)_3 
\right] \label{xxx}
\end{eqnarray}
This implies that in this case there are no FCNC in the up sector. 
From the requirement that $ \Delta_1  \Delta_2^\dagger  =0$, which is 
one of the conditions of Eq.~(\ref{gadel}), one concludes that Eq.~(\ref{xxx})
leads to the following three solutions for the diagonal matrices $d_1$, $d_2$:
\begin{subequations}
\begin{eqnarray}
\mbox{(up)} \quad d_1  =  \mbox{diag.} \left[\begin{array}{ccc}  
0 & \times  & \times 
\end{array}\right]; \qquad
d_2  =  \left[\begin{array}{ccc}  
\times & 0  & 0  
\end{array}\right] \label{22b}  \\
\mbox{(charm)} \quad d_1  =  \mbox{diag.} \left[\begin{array}{ccc}  
\times  & 0 & \times 
\end{array}\right]; \qquad
d_2  =  \left[\begin{array}{ccc}  
0 & \times & 0
\end{array}\right]  \label{22c} \\
\mbox{(top)} \quad d_1  =  \mbox{diag.} \left[\begin{array}{ccc}  
\times  & \times & 0 
\end{array}\right]; \qquad
d_2  =  \left[\begin{array}{ccc}  
0  & 0 & \times 
\end{array}\right] \label{22a} 
\end{eqnarray}
\end{subequations}
We have not included above, solutions  corresponding to 
the interchange of $d_1$, $d_2$. Without loss of generality, we shall
concentrate on one of the models, namely the ``top model". This
is the variant of the BGL models which satisfies all the constraints
of the MFV hypothesis. It is also the most interesting version, from
the phenomenological point of view, with strong natural suppression of FCNC
in $\Delta S =2$ transitions. In the top model, the $\Delta_i$ matrices have the
following form in any arbitrary WB:
\begin{eqnarray}
 \Delta_1 =  W_L^\dagger  \left[\begin{array}{ccc}  
(d_1)_1  & 0 & 0 \\
0  & (d_1)_2 &  0 \\
0 & 0 & 0 
\end{array}\right] W_R^u\\
 \Delta_2 =  W_L^\dagger  \left[\begin{array}{ccc}  
0  & 0 & 0 \\
0  & 0 &  0 \\
0 & 0 & (d_1)_3
\end{array}\right]  W_R^u
\end{eqnarray}
This implies that there are indeed  WBs where these matrices can be cast in the form
given by Eq.~(\ref{del}). This is obtained by choosing  unitary matrices
$W_L$ and $ W_R^u$ of the block form:
\begin{eqnarray}
W_L = \left[\begin{array}{cc}
(W_L)_{2\times2} & 0 \\
0 & e^{i \alpha} \end{array}\right] ; \quad 
 W_R^u = \left[\begin{array}{cc}
(W_R^u)_{2\times2} & 0 \\
0 & e^{i \beta} \end{array}\right]
 \label{wwr} 
\end{eqnarray}
leading to:
\begin{eqnarray}
 \Delta_1  =  W_L^\dagger  \left[\begin{array}{ccc}  
(d_1)_1  & 0 & 0 \\
0  & (d_1)_2 &  0 \\
0 & 0 & 0 
\end{array}\right] W_R^u =
\left[\begin{array}{ccc}  
\times  & \times & 0 \\
\times & \times &  0 \\
0 & 0 & 0 
\end{array}\right]; 
\label{usual1}\\
 \Delta_2   =  W_L^\dagger  \left[\begin{array}{ccc}  
0  & 0 & 0 \\
0  & 0 &  0 \\
0 & 0 & (d_1)_3
\end{array}\right]  W_R^u
=  \left[\begin{array}{ccc}  
0  & 0 & 0 \\
0 & 0 &  0 \\
0 & 0 & \times
\end{array}\right] 
\label{usual2}
\end{eqnarray}
Let us now see how the other conditions restrict the form of $\Gamma_1$
in this WB. In the top model,  the condition 
$\Gamma_1^\dagger  \Delta_2 =0$, leads to:
\begin{eqnarray}
\Gamma_1^\dagger  \Delta_2 =
\left[\begin{array}{ccc}  
\times  & \times & a \\
\times & \times &  b \\
\times & \times & c
\end{array}\right]
 \left[\begin{array}{ccc}  
0  & 0 & 0 \\
0 & 0 &  0 \\
0 & 0 & \times
\end{array}\right] =0
\label{ahah}
\end{eqnarray}
From  Eq.(\ref{ahah}) one obtains $a=b=c=0$, so $ \Gamma_1$ has the form
\begin{eqnarray}
\Gamma_1 = \left[\begin{array}{ccc}  
\times  & \times & \times \\
\times & \times &  \times\\
0 & 0 & 0
\end{array}\right]
\end{eqnarray}
in the basis where $\Delta_2$ has the form of  Eq.~(\ref{usual2}). The other condition
in Eqs.~(\ref{gadel}) requires $\Gamma_2^\dagger  \Delta_1 =0$ which
leads to:
\begin{eqnarray}
\Gamma_2^\dagger  \Delta_1 =
\left[\begin{array}{ccc}  
\alpha_1  & \alpha_2 & \alpha_3 \\
\beta_1 & \beta_2 &  \beta_3 \\
\gamma_1 & \gamma_2 & \gamma_3
\end{array}\right]
 \left[\begin{array}{ccc}  
A  & B & 0 \\
C & D &  0 \\
0 & 0 & 0
\end{array}\right] =0
\label{eheh}
\end{eqnarray}
From this equation one obtains:
\begin{eqnarray}
\alpha_1 A + \alpha_2 C = 0 \nonumber \\
\alpha_1 B + \alpha_2 D = 0
\label{ABCD}
\end{eqnarray}
Note that since, in the chosen WB, the up and charm quark only receive
mass from $\Delta_1$ the non-vanishing of $m_u$ and $m_c$ imply
$AD-BC \neq 0$ which together with Eqs.~(\ref{ABCD}) leads to:
\begin{equation}
\alpha_1 = \alpha_2 = 0
\end{equation}
In an entirely analogous manner, one can show that $\beta_1$, $\beta_2$
and $\gamma_1$, $\gamma_2$ vanish. One concludes then
that $\Gamma_2$ has the form:
\begin{eqnarray}
\Gamma_2   =  \left[\begin{array}{ccc}  
0 & 0 & 0  \\
0 & 0 & 0 \\
\times & \times &  \times 
\end{array}\right] 
\end{eqnarray}
This completes the proof that the relations of Eq.~(\ref{gadel}) are necessary 
and sufficient  conditions to have a BGL type model, with HFCNC in the 
down quark sector. 

Similarly, using the up and charm solutions of  Eqs.~(\ref{22b})  and (\ref{22c})
one obtains the other two BGL models with FCNC in the down sector.
The necessary and suffient conditions for BGL models with FCNC in the up sector 
can be written like those of  Eq.~(\ref{gadel}) with the r\^ ole of the
Yukawa matrices $ \Gamma_i$ and $ \Delta_i$  interchanged. 
These conditions are WB independent
and therefore they allow one to identify  BGL type models  when written in an
arbitrary WB where the zero texture patterns of the WB chosen by the symmetry 
are not present.\\

\noindent
{\bf \underline {The lowest invariants in the MFV framework and CP violation}} \\

It is instructive to evaluate the lowest non-trivial invariants in the case of BGL models.
In the general 2HDM one has:
\begin{eqnarray}
M_d N_d^{0 \dagger} = 
\frac{1}{2} \left[ v_1 v_2 \left( \Gamma_1  \Gamma_1^\dagger -
\Gamma_2 \Gamma_2^\dagger \right) + 
\left( v_2^2 \Gamma_2  \Gamma_1^\dagger -  v_1^2 \Gamma_1  \Gamma_2^\dagger
\right) \cos \theta  + \right. \nonumber  \\
+\left.  i \left( v_2^2 \Gamma_2  \Gamma_1^\dagger +  v_1^2 \Gamma_1  \Gamma_2^\dagger \right) \sin \theta \right]
\label{mmnn}
\end{eqnarray} 
It can be readily verified that in BGL models one has:
\begin{equation}
\mbox{tr}\left[ \Gamma_1  \Gamma_2^\dagger\right] = 0
\end{equation}
so that we obtain from Eq.~(\ref{mmnn}):
\begin{eqnarray}
I_1 \equiv \mbox{tr} (M_d N_d^{0 \dagger})  = \frac{1}{2} \mbox{tr}
\left[ v_1 v_2 \left( \Gamma_1  \Gamma_1^\dagger -
\Gamma_2 \Gamma_2^\dagger \right) \right]
\end{eqnarray}
The important point is that in BGL models $ M_d N_d^{0 \dagger} $ 
is an Hermitian matrix and thus:
\begin{eqnarray}
 \mbox{Im tr} (M_d N_d^{0 \dagger})  = 0
\label{dagger}
\end{eqnarray}
 From Eqs.~(\ref{nnn}) and (\ref{dagger}), it follows that in this class of models
Im$(N_d)_{jj} = 0$, thus avoiding too large e.d.m. for down-type quarks.

In order to extend the discussion of BGL type models to higher order WB
invariants it is instructive to review the formulation of these models in a more
generic way. Here we present some relations which greatly simplify the explicit
computation of higher order invariants in terms of physical quantities. In 
Ref.~\cite{Botella:2009pq} the special characteristics of BGL type models
were analysed and generalized. It was pointed out that the particular BGL 
example given explicitly at the beginning of this section, corresponds 
to a class of models where $N_d^0$ and $N_u^0$ can be writen as:
\begin{eqnarray}
N^0_d = \frac{v_2}{v_1} M_d - \left( \frac{v_2}{v_1} +  
\frac{v_1}{v_2}\right)   \mathcal{P}_{i}^{\gamma } \ M_d 
\label{bgl1} \\
N^0_u = \frac{v_2}{v_1} M_u - 
\left( \frac{v_2}{v_1} +  \frac{v_1}{v_2}\right)  
\mathcal{P}_{j}^{\beta} \ M_u \label{bgl2}
\end{eqnarray}
where 
$\mathcal{P} _{i}^{\alpha  }$ are the projection operators defined \cite{Botella:2004ks} by
\begin{eqnarray}
\mathcal{P}_{i}^{\alpha}=U_{\alpha L} P_{i} U_{\alpha L}^{\dagger }  
\label{projectors1} \\
\left(  P_{i} \right)_{lk}=\delta _{il}\delta _{ik}
\label{pilk}
\end{eqnarray}
and $\alpha$, $\beta$, $\gamma$ denote u (up) or d (down). BGL models
have $\gamma = \beta$ and therefore lead to HFCNC in one sector only. 
In BGL models we also have $i=j$. For  $\gamma = \beta = u$ there are 
HFCNC only in the down sector and vice versa for $\gamma = \beta = d$.
The example given at the beginning of this section corresponds to
 $\gamma = \beta = u$ and   $i=j=3$ and was presented in a particular
weak basis.That weak basis was chosen by the symmetry imposed on
the Lagrangian. Notice that the formulation presented here corresponds
to the generalization of the model to any weak basis. The choice $i=3$
together with $\gamma = \beta =u $ insures that the HFCNC are suppressed
by the third row of $V_{CKM}$. In the WB where $M_d$ is real and diagonal
this particular example corresponds to:
\begin{equation}
V_{CKM} \equiv U^\dagger_{uL}  U_{dL} = U^\dagger_{uL} 
\end{equation}
which leads to:
\begin{eqnarray}
M_d = D_d ,\qquad M_u = V_{CKM}^\dagger D_u  U^\dagger_{uR } \label{nnnn}\\
N^0_d \equiv N_d  = \frac{v_2}{v_1} D_d - \left( \frac{v_2}{v_1} +  
\frac{v_1}{v_2}\right)   V_{CKM}^\dagger P_3 V_{CKM} \ D_d 
\label{mbgl1} \\
N^0_u = \frac{v_2}{v_1} V_{CKM}^\dagger D_u  U^\dagger_{uR } - 
\left( \frac{v_2}{v_1} +  \frac{v_1}{v_2}\right)  
 V_{CKM}^\dagger P_3 D_u U^\dagger_{uR }  \label{mbgl2}
\end{eqnarray}
Eqs.~(\ref{bgl1}) -- (\ref{pilk}) together with the definition of $V_{CKM}$ enable
us to express all WB invariants in terms of physical quantities.
All six cases with $\gamma = \beta$ and $i=j$ can be obtained as the result
of a discrete symmetry \cite{Branco:1996bq}.

In Ref.~\cite{Botella:2009pq} a MFV expansion for $N^0_d$, $N^0_u$ with proper
transformation properties under a WB transformation, corresponding to a 
generalization of Eqs.~(\ref{bgl1}) and (\ref{bgl2}) is given by:
\begin{eqnarray}
N^0_d = \lambda_1 \ M_d + \lambda_{2i} \  U_{dL}P_i U^\dagger_{dL} \ M_d +
\lambda_{3i} \  U_{uL}P_i U^\dagger_{uL} \ M_d + ... \label{exp1} \\
N^0_u =  \tau_1 \ M_u + \tau_{2i} \  U_{uL}P_i U^\dagger_{uL} \ M_u +
\tau_{3i} \  U_{dL}P_i U^\dagger_{dL} \ M_u + ... \label{exp2}
\end{eqnarray}
In the quark mass eigenstate basis $N^0_d$, $N^0_u$  become:
\begin{eqnarray}
N_d = \lambda_1 \ D_d + \lambda_{2i} \ P_i  \ D_d + 
\lambda_{3i} \ (V_{CKM})^\dagger \ P_i \ V_{CKM} \ D_d + ...  
\label{abcd} \\
N_u =  \tau_1 \ D_u + \tau_{2i} \ P_i  \ D_u  + 
\tau_{3i} \ V_{CKM} \  P_i \ (V_{CKM})^\dagger \  D_u  + ... \label{ghjk} 
\end{eqnarray}
conforming explicitly with the requirement of depending only on the 
$V_{CKM}$ matrix.  This expansion contains as
particular cases the six BGL models mentioned above. Only these six models
can be obtained by means of an Abelian symmetry of the Lagrangian 
\cite{Ferreira:2010ir}, \cite{Botella:2011ne}. The symmetry also fixes the
coefficients of the expansion in the form given by Eqs.~(\ref{bgl1}) and (\ref{bgl2}).
The expansion given by Eqs.~(\ref{exp1}) and (\ref{exp2}) differs from the
usual one considered in the literature \cite{D'Ambrosio:2002ex}
by splitting each component of $M_d M^\dagger_d$
and $M_u M^\dagger_u$  into \cite{Botella:2004ks}:
\begin{equation}
H_\alpha = \sum_i {m^2_{\alpha}}_i\mathcal{P}^{\alpha}_{i}
\end{equation}
and allowing for different coefficients for each  term of the expansion 
in $\mathcal{P}^{\alpha}_{i}$ with a different index $i$.
In this sense the expansion given here is more general and contains the one used
in the literature by many authors as a special case.

It is well known that in the SM the lowest order WB invariant sensitive to
CP violation is given by Eq.~(\ref{eq31}) and has dimension twelve
in powers of mass. Obviously, this invariant is also relevant for BGL type models.
However, in BGL type models we have a richer flavour structure parametrized 
in terms of the four matrices $M_d$, $M_u$, $N^0_d$ and $N^0_u$ rather
than the two mass matrices of the SM. As a result, in this case the 
lowest order invariant sensitive to CP violation is of lower order, namely:
 \begin{equation}
I_9^{CP} \equiv \mbox{Im tr} \left[ M_d  N_d^{0 \dagger} M_d M^\dagger_d
M_u M^\dagger_u M_d M^\dagger_d \right] 
\end{equation}
In BGL models invariants that see CP violation must
contain flavour matrices both from the up and down sector. In fact
the sector that has HFCNC has couplings that are proportional to only one 
row of $V_{CKM}$ and it is always possible to choose  a parametrization where
any single row of $V_{CKM}$ is real. 
This invariant can be readily evaluated using Eqs.~(\ref{nnnn}), (\ref{mbgl1}),
which correspond to the specific BGL model given at the beginning of this section
with $\gamma = u$ and $i=3$, and one obtains:
\begin{eqnarray}
I_9^{CP} ( \gamma =u, i=3) &  &= 
 -\left( \frac{v_2}{v_1} +  \frac{v_1}{v_2}\right)  (m^2_b - m^2_s)
(m^2_b - m^2_d) (m^2_s - m^2_d)  \times \nonumber \\
& &  \times (m^2_c - m^2_u) \mbox{Im} \left( V^\ast_{22}  V_{32} V^\ast_{33}  V_{23} \right)
\end{eqnarray}
This result is in agreement with the MFV character of BGL models namely,
all flavour changing and CP violation are controlled by $V_{CKM}$, therefore 
this CP violating quantity must be proportional to the imaginary part of
rephasing invariant quartets of  $V_{CKM}$ as in the SM \cite{Branco:1999fs}.
Another important result is that $I_9^{CP} ( \gamma =u, i=3) $ is different 
from zero even if $m_t = m_c$  or $m_t = m_u$. In fact the discrete symmetry
leading to this specific BGL model singles out the top quark 
\cite{Branco:1996bq}. It is important
to emphazise that this invariant is defined in such a way that the trace involves
the sum over all quarks, therefore it can be related to the baryon asymmetry 
generated at the electroweak phase transition \cite{Gavela:1993ts}, \cite{Gavela:1994ds},
\cite{Gavela:1994dt}, \cite{Hou:2008xd}. 

In the BGL model defined by
$\gamma =d$, $i=1$, where:
\begin{eqnarray}
N^0_d = \frac{v_2}{v_1} M_d - \left( \frac{v_2}{v_1} +  
\frac{v_1}{v_2}\right)   \mathcal{P}_{1}^d \ M_d 
\label{bgl33} \\
N^0_u = \frac{v_2}{v_1} M_u - 
\left( \frac{v_2}{v_1} +  \frac{v_1}{v_2}\right)  
\mathcal{P}_{1}^d \ M_u \label{bgl22}
\end{eqnarray}
we can get an enhancement in the CP violating contribution to the
baryon asymmetry of the order:
\begin{eqnarray}
\frac{ I_9^{CP} ( \gamma =d, i=1) }{I_1^{CP}} \ \frac{E^{12}} {E^{8}} 
\simeq
\left( \frac{v_2}{v_1} +  \frac{v_1}{v_2}\right) \frac {E^{4}}{m_b^2 m_s^2}
\end{eqnarray}
where $E$ is the scale relevant for baryogenesis at the electroweak phase 
transition. For $E \sim 100$GeV we get an enhancement of about $10^{10}$.
This enhancement can be traced to the fact that this model singles out the $d$
quark in such a way that  the only mass difference involving down quarks
appearing in  $I_9^{CP} ( \gamma =d, i=1) $ is the suppression term
$(m^2_b - m^2_s)$  unlike in $I_1^{CP}$  where the three different down square mass 
differences appear, so that the ratio of this invariant by $I_1^{CP}$ is
larger by a factor of the order $ E^4 / (m^2_b - m^2_d)(m^2_s - m^2_d)$ \\

It is instructive to make use of Eqs.~(\ref{nnnn}), (\ref{mbgl1}) to compute
$I_2$ which is real in this case:
\begin{eqnarray}
I_2 = 
 -2 m_d^2 m_s^2 (m_s^2 - m_d^2)^2  \left( \frac{v_2}{v_1} +  
\frac{v_1}{v_2}\right)^2 |V_{31}|^2 |V_{32}|^2 - \nonumber \\
- 2 m_d^2 m_b^2 (m_b^2 - m_d^2)^2  \left( \frac{v_2}{v_1} +  
\frac{v_1}{v_2}\right)^2 |V_{33}|^2 |V_{31}|^2 - \label{qual} \\
- 2 m_s^2 m_b^2 (m_b^2 - m_s^2)^2 \left( \frac{v_2}{v_1} +  
\frac{v_1}{v_2}\right)^2 |V_{33}|^2 |V_{32}|^2\nonumber
\end{eqnarray}
the dominant term is the last one.

\subsection {Two Higgs doublets with the NNI texture}

Some time ago \cite{Branco:1988iq} it has been shown that, in the three generation
SM, starting with arbitrary Yukawa couplings, one can always make a
WB transformation such that the quark mass matrices $M_d$, $M_u$ get the form:
\begin{eqnarray}
 M_d & = & \left[\begin{array}{ccc}  
0  & a_d & 0 \\
a^\prime_d & 0 &  b_d \\
0 & b^\prime_d & c_d
\end{array}\right]; \qquad
M_u  =  \left[\begin{array}{ccc}  
0  & a_u & 0 \\
a^\prime_u & 0 &  b_u \\
0 & b^\prime_u & c_u
\end{array}\right]
\label{nni}
\end{eqnarray}
this form, usually denoted nearest-neighbour-interaction (NNI) basis has
no physical implications in the context of the SM with one Higgs doublet.
If one further assumes that $M_d$, $M_u$ are Hermitian in the NNI basis
(i.e., $a^\prime_{d(u)}  = a^\ast_{d(u)}$, $b^\prime_{d(u)} = b^\ast_{d(u)}$) then one
obtains the Fritzsch Ansatz \cite{Fritzsch:1977vd} which does have physical implications,
correctly predicting $|V_{us}|$ but making a wrong prediction for $|V_{cb}|$,
taking into account that $m_t \gg m_c$. This implies that the original
Fritzsch Ansatz has been ruled out. Recently, it has been shown that one 
can reproduce all the current data on quark masses and mixing, by 
allowing deviations of Hermiticity of about $20\%$ in the NNI form. It was 
also shown \cite{Branco:2010tx} that, in the context of 2HDM,  one can 
obtain the NNI form for the quark mass matrices, through the introduction of
a $Z_4$ symmetry in the Lagrangian, which leads to:
\begin{eqnarray}
 \frac{v_1}{\sqrt 2}\Gamma_1 & = & \left[\begin{array}{ccc}  
0  & a_d & 0 \\
a^\prime_d & 0 &  0 \\
0 & 0 & c_d
\end{array}\right]; \qquad
  \frac{v_2 e^{i \theta}}{\sqrt 2}\Gamma_2   =  \left[\begin{array}{ccc}  
0  & 0 & 0 \\
0  & 0 &  b_d \\
0 & b^\prime_d & 0
\end{array}\right]  \label{nnid} \\
  \frac{v_1}{\sqrt 2}\Delta_1  & = & \left[\begin{array}{ccc}  
0  & 0 & 0 \\
0 & 0 &  b_u \\
0 & b^\prime_u & 0
\end{array}\right]; \qquad 
   \frac{v_2 e^{-i \theta}}{\sqrt 2}\Delta_2   =  \left[\begin{array}{ccc}  
0  & a_u & 0 \\
a^\prime_u & 0 &  0 \\
0 & 0 & c_u
\end{array}\right]  \label{nniu}
\end{eqnarray}
It is clear that the couplings of Eqs.~(\ref{nnid}), (\ref{nniu}) lead to HFCNC
in both the up and down sectors. 
In this section, we evaluate some of the previously defined WB invariants,
illustrating their usefulness in the analysis of HFCNC and CP violating effects. \\

Let us consider $I_1$ again. It can be easily checked that this invariant
is real in the NNI case.:
\begin{equation}
I_1 \equiv \mbox{tr} (M_d N_d^{0 \dagger} )= \frac{v_2}{v_1} ( a_d  a_d^\ast + 
a^\prime_d {a^\prime_d}^\ast) 
-  \frac{v_1}{v_2} ( b_d  b_d^\ast +  b^\prime_d {b^\prime_d}^\ast) +
 \frac{v_2}{v_1}  c_d c_d^\ast
\end{equation}

The same is true for $I_2$ which in this case is given by: \\
\begin{eqnarray}
I_2 \equiv \mbox{tr} \left[ M_d N_d^{0 \dagger}, M_d M_d^ \dagger \right]^2 =
\left( \frac{v_2}{v_1} +  \frac{v_1}{v_2}\right)^2 \ 2 \  c_d c_d^\ast 
\left[ c_d c_d^\ast  b_d  b_d^\ast  b^\prime_d {b^\prime_d}^\ast  \right. + \nonumber \\
\left.  + (b^\prime_d {b^\prime_d}^\ast)^2 a_d  a_d^\ast  + (b_d  b_d^\ast)^2 a^\prime_d {a^\prime_d}^\ast - a^\prime_d {a^\prime_d}^\ast b_d  b_d^\ast b^\prime_d 
{b^\prime_d}^\ast
-  a_d  a_d^\ast  b_d  b_d^\ast b^\prime_d {b^\prime_d}^\ast
\right]
\end{eqnarray}
In order to compare this result to the one obtained in the MFV case given by Eq.~(\ref{qual})
we rewrite the coefficients of the NNI mass matrices in terms of quark masses
using the approximate relations of Ref.~\cite{Branco:2010tx}:
\begin{equation}
c_d c_d^\ast \sim m_b^2, \ \ |a_d | \sim  | a^\prime_d | \sim \sqrt{m_d m_s}, \ \ 
|b_d | \sim  | b^\prime_d | \sim \sqrt{m_s m_b}
\end{equation}
which lead to:
\begin{equation}
I_2  \sim 2 \left( \frac{v_2}{v_1} +  \frac{v_1}{v_2}\right)^2 
m_b^6 m_s^2 
\end{equation}
There are  similarities between the dominant term in the MFV case 
and  the NNI case, but in the NNI case there is no suppression factor given by the 
$V_{CKM}$ matrix elements. Therefore HFCNC are potentially more dangerous in  NNI 
models than in the MFV case.  Another important point is the fact that 
in the NNI case the lowest invariants in powers of masses, sensitive to CP violation, 
are much lower than $I_9^{CP}$. One such example is:  
\begin{equation}
\mbox{Im tr} \left[ M_d  N_d^{0 \dagger} 
M_u M^\dagger_u  \right]  \sim \left( \frac{v_2}{v_1} +  \frac{v_1}{v_2}\right)
m_c^{\frac{1}{2}} m_t^{\frac{3}{2}} m_s^{\frac{1}{2}} m_b^{\frac{3}{2}} \sin \beta
\end{equation}
where the angle $\beta$ is one of the two factorizable phases that cannot
be removed from the mass matrices by  rephasing of the quark fields.
Note that in the NNI case it is possible to choose a WB where $M_d$ (or else
$M_u$)  is real by rephasing quark fields. In this WB $N^0_d$ (or else $N^0_u$)
is also real. Further rephasing on the righthanded side allows to remove
three phases from the other mass matrix and also, at the same time, from the
corresponding $N^0$ matrix, so that we are left with only two meaningful 
factorizable phases in the other mass matrix coinciding with the two phases
left in the corresponding $N^0$ matrix.
In Ref.~\cite{Branco:2010tx}  these two phases are evaluated, their sine
is roughly of order one.  Implications for the baryon asymmetry of the 
Universe are also important in this case.

\section{Conclusions}
We have presented a discussion of various flavour aspects of the general 
2HDM. In particular, we have shown that flavour mixing in the 2HDM
can be parametrized by a set of unitary matrices which arise from the 
misalignment in flavour space of various pairs of Hermitian matrices
constructed from the Yukawa couplings of the 2HDM. These unitary 
mixing matrices are entirely analogous to the CKM matrix which arises 
in the SM from the misalignment of $H_d \equiv M_d M^\dagger_d$
and $H_u \equiv M_u M^\dagger_u$. Some of the CP violating phases are 
entirely analogous to the CKM phase, reflecting the non-vanishing of the
imaginary parts of the various invariant quartets of the above unitary 
flavour matrices which arise in the 2HDM. We construct various WB 
invariants which can play a crucial r\^ ole in the analysis of both CP 
violation and FCNC.  Apart from a general analysis, we also applied the WB 
invariants to the study of specific flavour models,  in the framework of 2HDM, 
such as MFV models of BGL type and models with a NNI structure.
It is likely that the flavour structure of the 2HDM is not generic, reflecting 
on the contrary, the presence of some flavour symmetry. The WB invariants
which we have constructed can be very useful in the study of new sources
of CP violation in 2HDM constrained by some flavour symmetry. In particular,
these WB invariants can be applied to the study of Higgs mediated FCNC
in flavoured 2HDM. We also point out that
in the 2HDM with MFV as well as in NNI models, CP-odd WB invariants 
arise in terms of much lower
powers of masses than the CP-odd invariant of the SM, a feature which
can have important implications for baryogenesis.
The recent discovery of a Higgs boson at the LHC is an important step
towards understanding the electroweak symmetry breaking
sector. The LHC and, in the future, a linear collider
will play an important r\^ ole  in putting further constraints on 
different two Higgs doublet model scenarios \cite{review},  \cite{Jung:2010ik}, \cite{Ferreira:2011aa}
taking into account, in particular, the distinguishing features between models with 
NFC and with MFV \cite{Krawczyk:2008zz}, \cite{Buras:2010mh}, \cite{Cervero:2012cx}, \cite{Mader:2012pm}, \cite{Basso:2012st}, \cite{Altmannshofer:2012ar}.

\section*{Acknowledgements}
This work was partially supported by Funda\c c\~ ao para a Ci\^ encia e
a Tecnologia (FCT, Portugal) through the projects CERN/FP/123580/2011
PTDC/FIS/ 098188/2008 and CFTP-FCT Unit 777 which are partially funded 
through POCTI (FEDER), 
by Accion Complementaria Luso-Espanhola AIC-D-2011-0809, by European FEDER,
Spanish MINECO under grant FPA2011--23596, by GVPROMETEO 2010--056 and
by Marie Curie Initial Training Network "UNILHC" PITN-GA-2009-237920.
MNR is grateful to the Theory Division of CERN where her present  work was done
as CERN Scientific Associate. FB and GCB are also grateful to CERN for hospitality
during their visits. The authors visited each other's Institutes during the
preparation of this work and each time were warmly welcomed.


\begin{thebibliography}{999}

\bibitem{review}
For recent reviews see:
A.~Djouadi,
  Eur.\ Phys.\ J.\ C {\bf 59} (2009) 389
  [arXiv:0810.2439 [hep-ph]];
G.~C.~Branco, P.~M.~Ferreira, L.~Lavoura, M.~N.~Rebelo, M.~Sher and J.~P.~Silva,
  Phys.\ Rept.\  {\bf 516} (2012) 1
  [arXiv:1106.0034 [hep-ph]].
See also  J.~F.~Gunion, H.~E.~Haber, G.~L.~Kane and S.~Dawson,
  ``The Higgs Hunter's Guide,''
  Front.\ Phys.\  {\bf 80} (2000) 1;
  

\bibitem{Lee:1973iz}
  T.~D.~Lee,
  Phys.\ Rev.\ D {\bf 8} (1973) 1226.

\bibitem{Branco:1999fs}
  G.~C.~Branco, L.~Lavoura and J.~P.~Silva,
  Int.\ Ser.\ Monogr.\ Phys.\  {\bf 103} (1999) 1.

\bibitem{Glashow:1976nt}
  S.~L.~Glashow and S.~Weinberg,
  Phys.\ Rev.\ D {\bf 15} (1977) 1958.

\bibitem{Branco:1980sz}
  G.~C.~Branco,
  Phys.\ Rev.\ D {\bf 22} (1980) 2901.

\bibitem{Weinberg:1976hu}
  S.~Weinberg,
  Phys.\ Rev.\ Lett.\  {\bf 37} (1976) 657.

\bibitem{Branco:1985aq}
  G.~C.~Branco and M.~N.~Rebelo,
  Phys.\ Lett.\ B {\bf 160} (1985) 117.

\bibitem{Hamzaoui:1998nu}
  C.~Hamzaoui, M.~Pospelov and M.~Toharia,
  Phys.\ Rev.\ D {\bf 59} (1999) 095005
  [hep-ph/9807350].

\bibitem{Babu:1999hn}
  K.~S.~Babu and C.~F.~Kolda,
  Phys.\ Rev.\ Lett.\  {\bf 84} (2000) 228
  [hep-ph/9909476].

\bibitem{Isidori:2001fv}
  G.~Isidori and A.~Retico,
  JHEP {\bf 0111} (2001) 001
  [hep-ph/0110121].

\bibitem{Branco:1996bq}
  G.~C.~Branco, W.~Grimus and L.~Lavoura,
  Phys.\ Lett.\  B {\bf 380} (1996) 119
  [arXiv:hep-ph/9601383].

\bibitem{Botella:2009pq}
  F.~J.~Botella, G.~C.~Branco and M.~N.~Rebelo,
  Phys.\ Lett.\ B {\bf 687} (2010) 194
  [arXiv:0911.1753 [hep-ph]].

\bibitem{D'Ambrosio:2002ex}
  G.~D'Ambrosio, G.~F.~Giudice, G.~Isidori and A.~Strumia,
  Nucl.\ Phys.\ B {\bf 645} (2002) 155
  [hep-ph/0207036].

\bibitem{Chivukula:1987py}
  R.~S.~Chivukula and H.~Georgi,
  Phys.\ Lett.\ B {\bf 188} (1987) 99.

\bibitem{Hall:1990ac}
  L.~J.~Hall and L.~Randall,
  Phys.\ Rev.\ Lett.\  {\bf 65} (1990) 2939.

\bibitem{Buras:2000dm}
  A.~J.~Buras, P.~Gambino, M.~Gorbahn, S.~Jager and L.~Silvestrini,
  Phys.\ Lett.\ B {\bf 500} (2001) 161
  [hep-ph/0007085].

\bibitem{Branco:2010tx}
  G.~C.~Branco, D.~Emmanuel-Costa and C.~Simoes,
  Phys.\ Lett.\ B {\bf 690} (2010) 62
  [arXiv:1001.5065 [hep-ph]].

\bibitem{Branco:1987mj}
  G.~C.~Branco and L.~Lavoura,
  Phys.\ Lett.\ B {\bf 208} (1988) 123.

\bibitem{Bernabeu:1986fc}
  J.~Bernabeu, G.~C.~Branco and M.~Gronau,
  Phys.\ Lett.\  B {\bf 169} (1986) 243.

\bibitem{Jarlskog:1985ht}
  C.~Jarlskog,
  Phys.\ Rev.\ Lett.\  {\bf 55} (1985) 1039.

\bibitem{Branco:2005em}
  G.~C.~Branco, M.~N.~Rebelo and J.~I.~Silva-Marcos,
  Phys.\ Lett.\ B {\bf 614} (2005) 187
  [hep-ph/0502118].

\bibitem{Lavoura:1994fv}
  L.~Lavoura and J.~P.~Silva,
  Phys.\ Rev.\ D {\bf 50} (1994) 4619
  [hep-ph/9404276].

\bibitem{Botella:1994cs}
  F.~J.~Botella and J.~P.~Silva,
  Phys.\ Rev.\ D {\bf 51} (1995) 3870
  [hep-ph/9411288].

\bibitem{Dreiner:2007yz}
  H.~K.~Dreiner, J.~S.~Kim, O.~Lebedev and M.~Thormeier,
  Phys.\ Rev.\ D {\bf 76} (2007) 015006
  [hep-ph/0703074 [HEP-PH]].

\bibitem{Branco:2011aa}
  G.~C.~Branco and J.~I.~Silva-Marcos,
  arXiv:1112.1631 [hep-ph].

\bibitem{Jenkins:2009dy}
  E.~E.~Jenkins and A.~V.~Manohar,
  JHEP {\bf 0910} (2009) 094
  [arXiv:0907.4763 [hep-ph]].


\bibitem{Botella:2011ne}
  F.~J.~Botella, G.~C.~Branco, M.~Nebot and M.~N.~Rebelo,
  JHEP {\bf 1110} (2011) 037
  [arXiv:1102.0520 [hep-ph]].

\bibitem{Botella:2004ks}
  F.~J.~Botella, M.~Nebot and O.~Vives,
  JHEP {\bf 0601} (2006) 106
  [arXiv:hep-ph/0407349].

\bibitem{Ferreira:2010ir}
  P.~M.~Ferreira and J.~P.~Silva,
  Phys.\ Rev.\ D {\bf 83} (2011) 065026
  [arXiv:1012.2874 [hep-ph]].

\bibitem{Gavela:1993ts}
  M.~B.~Gavela, P.~Hernandez, J.~Orloff and O.~Pene,
  Mod.\ Phys.\ Lett.\ A {\bf 9} (1994) 795
  [hep-ph/9312215, hep-ph/9312215].

\bibitem{Gavela:1994ds}
  M.~B.~Gavela, M.~Lozano, J.~Orloff and O.~Pene,
  Nucl.\ Phys.\ B {\bf 430} (1994) 345
  [hep-ph/9406288].

\bibitem{Gavela:1994dt}
  M.~B.~Gavela, P.~Hernandez, J.~Orloff, O.~Pene and C.~Quimbay,
  Nucl.\ Phys.\ B {\bf 430} (1994) 382
  [hep-ph/9406289].

\bibitem{Hou:2008xd}
  W.~-S.~Hou,
  Chin.\ J.\ Phys.\  {\bf 47} (2009) 134
  [arXiv:0803.1234 [hep-ph]].


\bibitem{Branco:1988iq}
  G.~C.~Branco, L.~Lavoura and F.~Mota,
  Phys.\ Rev.\ D {\bf 39} (1989) 3443.

\bibitem{Fritzsch:1977vd}
  H.~Fritzsch,
  Phys.\ Lett.\ B {\bf 73} (1978) 317.

\bibitem{Jung:2010ik}
  M.~Jung, A.~Pich and P.~Tuzon,
  JHEP {\bf 1011} (2010) 003
  [arXiv:1006.0470 [hep-ph]].

\bibitem{Ferreira:2011aa}
  P.~M.~Ferreira, R.~Santos, M.~Sher and J.~P.~Silva,
  Phys.\ Rev.\ D {\bf 85} (2012) 077703
  [arXiv:1112.3277 [hep-ph]].

\bibitem{Krawczyk:2008zz}
  M.~Krawczyk,
  PoS CHARGED {\bf 2008} (2008) 017.

\bibitem{Buras:2010mh}
  A.~J.~Buras, M.~V.~Carlucci, S.~Gori and G.~Isidori,
  JHEP {\bf 1010} (2010) 009
  [arXiv:1005.5310 [hep-ph]].

\bibitem{Cervero:2012cx}
  E.~Cervero and J.~-M.~Gerard,
  Phys.\ Lett.\ B {\bf 712} (2012) 255
  [arXiv:1202.1973 [hep-ph]].

\bibitem{Mader:2012pm}
  W.~Mader, J.~-h.~Park, G.~M.~Pruna, D.~Stockinger and A.~Straessner,
  JHEP {\bf 1209} (2012) 125
  [arXiv:1205.2692 [hep-ph]].

\bibitem{Basso:2012st}
  L.~Basso, A.~Lipniacka, F.~Mahmoudi, S.~Moretti, P.~Osland, G.~M.~Pruna and M.~Purmohammadi,
  arXiv:1205.6569 [hep-ph].

\bibitem{Altmannshofer:2012ar}
  W.~Altmannshofer, S.~Gori and G.~D.~Kribs,
  arXiv:1210.2465 [hep-ph].


\end{thebibliography}
\end{document}